\documentclass[twocolumn,showpacs,superscriptaddress,amsmath,amssymb,nofootinbib]{revtex4-2}
\usepackage{graphicx}
\usepackage{epsfig}
\usepackage{overpic}
\usepackage{dcolumn}
\usepackage{ulem}
\usepackage{bm}
\usepackage{lineno}
\usepackage{xspace}
\usepackage{multirow}
\usepackage{epstopdf}
\usepackage{xcolor}
\usepackage{soul}
\usepackage[colorlinks,allcolors=black]{hyperref}
\usepackage{verbatim}
\usepackage{enumitem}
\usepackage{todonotes}
\usepackage{subfigure}
\usepackage{dcolumn}% Align table columns on decimal point
\usepackage{bm}% bold math
\usepackage{threeparttable}
\usepackage{multirow}
\usepackage[figuresright]{rotating}
\usepackage{makecell}
\usepackage{verbatim}% 注释
\usepackage{subfigure}
\include{def-com}

\newcommand{\mev}{\ensuremath{\;\text{MeV}\xspace}}

\newcommand{\gev}{\ensuremath{\;\text{GeV}\xspace}}

\begin{document}
\title{\boldmath The third peak structure in the double $J/\psi$ spectrum}

\author{Peng-Yu Niu} %\email{niupy@m.scnu.edu.cn}
\address{Guangdong Provincial Key Laboratory of Nuclear Science, Institute of Quantum Matter, South China Normal University, Guangzhou 510006, China}
\affiliation{Guangdong-Hong Kong Joint Laboratory of Quantum Matter, Southern Nuclear Science Computing Center, South China Normal University, Guangzhou 510006, China}

\author{Zhenyu Zhang}%\email{zhenyuzhang@m.scnu.edu.cn}
\address{Guangdong Provincial Key Laboratory of Nuclear Science, Institute of Quantum Matter, South China Normal University, Guangzhou 510006, China}
\affiliation{Guangdong-Hong Kong Joint Laboratory of Quantum Matter, Southern Nuclear Science Computing Center, South China Normal University, Guangzhou 510006, China}

\author{Qian Wang}\email{qianwang@m.scnu.edu.cn}
\address{Guangdong Provincial Key Laboratory of Nuclear Science, Institute of Quantum Matter, South China Normal University, Guangzhou 510006, China}
\affiliation{Guangdong-Hong Kong Joint Laboratory of Quantum Matter, Southern Nuclear Science Computing Center, South China Normal University, Guangzhou 510006, China}

\author{Meng-Lin Du}\email{du.ml@uestc.edu.cn}
\address{School of Physics, University of Electronic Science and Technology of China, Chengdu 611731, China}
%\affiliation{XX}

\date{\today}
%\date{\today}

%%%%%%%%%%%%%%%% ．，
\begin{abstract}
\end{abstract}
%\pacs{xx.Gp,  yy.Rj, zz.Dh}

\maketitle

Quantum Chromodynamics (QCD) is the fundamental theory of strong interaction, whose color confinement property
allows for the existence of any color-neutral objects, i.e. the so-called hadrons.
In the conventional quark model, hadrons are classified into mesons 
and baryons, made of quark-antiquark and three quarks, respectively. 
However, since the observation of the $\chi_{c1}(3872)$,
a.k.a. $X(3872)$, in 2003, numerous hadrons beyond the above two configurations
have been discovered, i.e. exotic hardons. On the other hand, as unique direct measurable 
objects, hadrons provide a way to reveal the mystery of the nonperturbative QCD. 
Therefore, continuous efforts have been put forward by both experimentalists and theorists
to understand the formation of hadrons, in particular exotic hadrons. 
Up to now, dozens of exotic candidates have been reported and most of them
are in the heavy quarkonium energy region. For instance,
the most famous hidden-charm pentaquarks $P_c(4312)^+$, $P_c(4440)^+$, $P_c(4457)^+$,
 the first doubly charmed tetraquark $T_{cc}^+$,
 the fully charmed tetraquark $X(6900)$ and so on.
  A considerable portion of them have nearby $S$-wave thresholds and
 can be viewed as either hadronic molecular candidates~\cite{Guo:2017jvc}
 or cusp effects~\cite{Guo:2019twa}.
  
Among those exotic candidates, the fully heavy systems are particularly interesting
due to the absence of light quarks. 
The LHCb~\cite{LHCb:2018uwm} and CMS~\cite{CMS:2020qwa} collaboration  performed a search for narrow resonances
 in the $\Upsilon(1S)\mu^+\mu^-$ channel around the four-bottom quark mass region. 
 Unfortunately, no significant excess of events was observed. 
 The situation was broken up by the
 LHCb collaboration~\cite{LHCb:2020bwg} in 2020, which reported 
 a narrow structure around $6.9~\mathrm{GeV}$ and a broad structure
 in the $J/\psi J/\psi$ channel. In
addition to the two structures claimed in the main text,
the LHCb collaboration, in the supplementary material,  also performed a fit with three 
Breit-Wigner (BW) lineshapes. Those three structures
were reported by the CMS collaboration two years later~\cite{CMS:2022yhl}. 
   Meanwhile, the ATLAS Collaboration also reported
   the existence of the $X(6900)$ in the $J/\psi \psi^\prime$ mass spectrum in 2022~\cite{ATLAS:2022hhx}. 
   Besides the mentioned peak structures,
 there is also a dip around $6.75~\mathrm{GeV}$ in the $J/\psi J/\psi$ invariant mass spectrum.
 The controversy about the structure around $7.2~\mathrm{GeV}$ among the three collaborations
 is the key to approaching the mystery of fully charmed tetraquarks. 
\begin{figure}[htbp]
\centering
\includegraphics[scale=0.3]{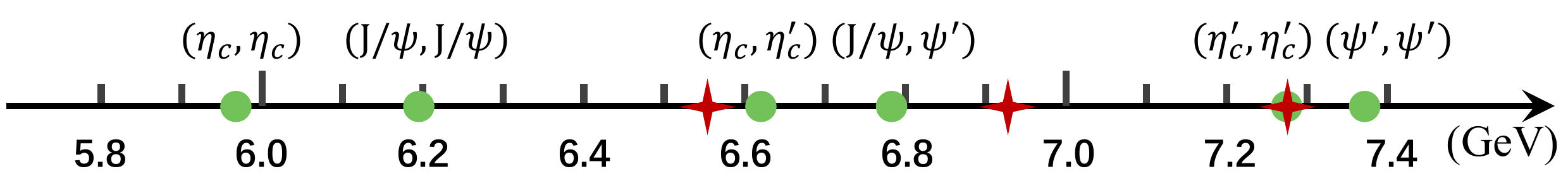}
\caption{(Color online) The positions of the double $S$-wave charmonia thresholds (green circles) below $7.4~\mathrm{GeV}$ comparing with
the three peak positions (red stars) from Model I in the CMS analysis~\cite{CMS:2022yhl}.}
\label{fig:threshold}
\end{figure}
It is noticed that the first and the third peaks of the CMS analysis~\cite{CMS:2022yhl}
are close to the $\eta_c\eta_c^\prime$ and $\eta_c^\prime\eta_c^\prime$, thresholds, respectively,
\footnote{ Here and what follows, we use $\eta_c^\prime$ and $\psi^\prime$
to denote the $\eta_c(2S)$ and $\psi(2S)$ charmonia, respectively.} as shown in Fig.~\ref{fig:threshold}~\footnote{ Although the $\Xi_{cc}^{++}\bar \Xi_{cc}^{++}$ ($7.24$ GeV) and $\Xi_{cc}^{+} \bar \Xi_{cc}^{+}$($7.04$ GeV) thresholds are also around $7.2$ GeV, their productions are suppressed with comparison to that of double charmonium. In addition, their couplings to di-$J/\psi$ are suppressed due to the annihilation of the light quark pair.}. 
Especially, the third peak is $12\pm 20~\mev$ 
  above the $\eta_c^\prime \eta_c^\prime$ threshold for the CMS analysis~\cite{CMS:2022yhl}
  and $55^{+36}_{-42}~\mev$
   below the $\eta_c^\prime \eta_c^\prime$ threshold for the ATLAS analysis~\cite{ATLAS:2022hhx}. 
   Considering the uncertainties, they coincide with the $\eta_c^\prime\eta_c^\prime$ threshold as shown in Fig.~\ref{fig:threshold}.
   Therefore, to shed light on the nature of the third peak structure
   one cannot avoid the effect of the $\eta_c^\prime \eta_c^\prime$ channel.
   In this short communication, we aim at exploring the nature of the third peak
   by considering the $S$-wave charmonium pair in $pp$ collision. 

In the hadronic molecular picture, the production of exotic hadrons involves two ingredients, i.e. the bare production vertex and the final-state interaction. As the production of excited charmonium is smaller than that of ground charmonium, we consider the scattering among the lowest two S-wave charmonium doublets (see the supplementary material). In addition, the double pseudo-scalar charmonia can couple to double vector charmonia
by rearranging the charm and anticharm quarks, which is allowed by the heavy quark spin symmetry.
In the heavy quark limit, the dynamics does not depend on the spin
of heavy quarks, i.e. the Heavy Quark Spin Symmetry (HQSS).
The $S$-wave potentials for
the scattering among $1S$ and $2S$ charmonia with
quantum numbers $0^{++}$, $2^{++}$ can be related via HQSS \footnote{ The HQSS breaking effect, which is considered as the higher order contribution,
for charm system is of the order $\Lambda_{\mathrm{QCD}}/m_c$ and neglected in the current calculation. Here $\Lambda_{\mathrm{QCD}}$ is 
the typical QCD nonperturbative momentum scale
 and $m_c$ is the charm quark mass.}.
 The production amplitude for each quantum number 
can be obtained by solving
Lippmann-Schwinger equation (LSE). The details can be found in the supplementary material.
To explore the underlying dynamics, two schemes, 
i.e. incoherence and coherence with the background contribution denoted as Scheme I and Scheme II, respectively,
are performed as a comparison. We start from the
$J^{PC}=0^{++}$ case with the $\eta_c\eta_c$, $J/\psi J/\psi$,
$\eta_c\eta_c^\prime$, $J/\psi \psi^\prime$, $\eta_c^\prime\eta_c^\prime$, $\psi^\prime\psi^\prime$
as the dynamical channels. 
\begin{figure*}[htbp]
\centering
\includegraphics[scale=0.4]{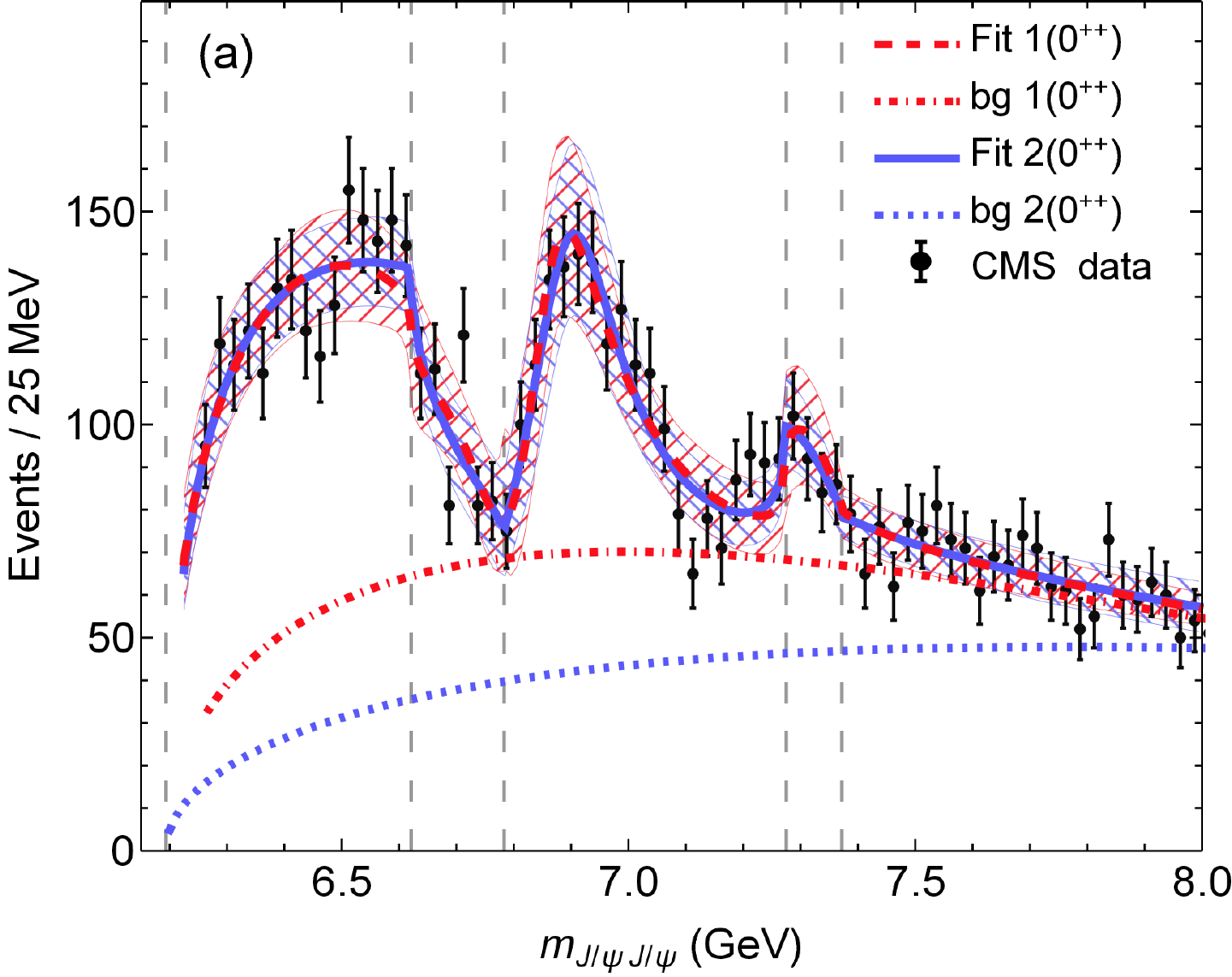}\includegraphics[scale=0.4]{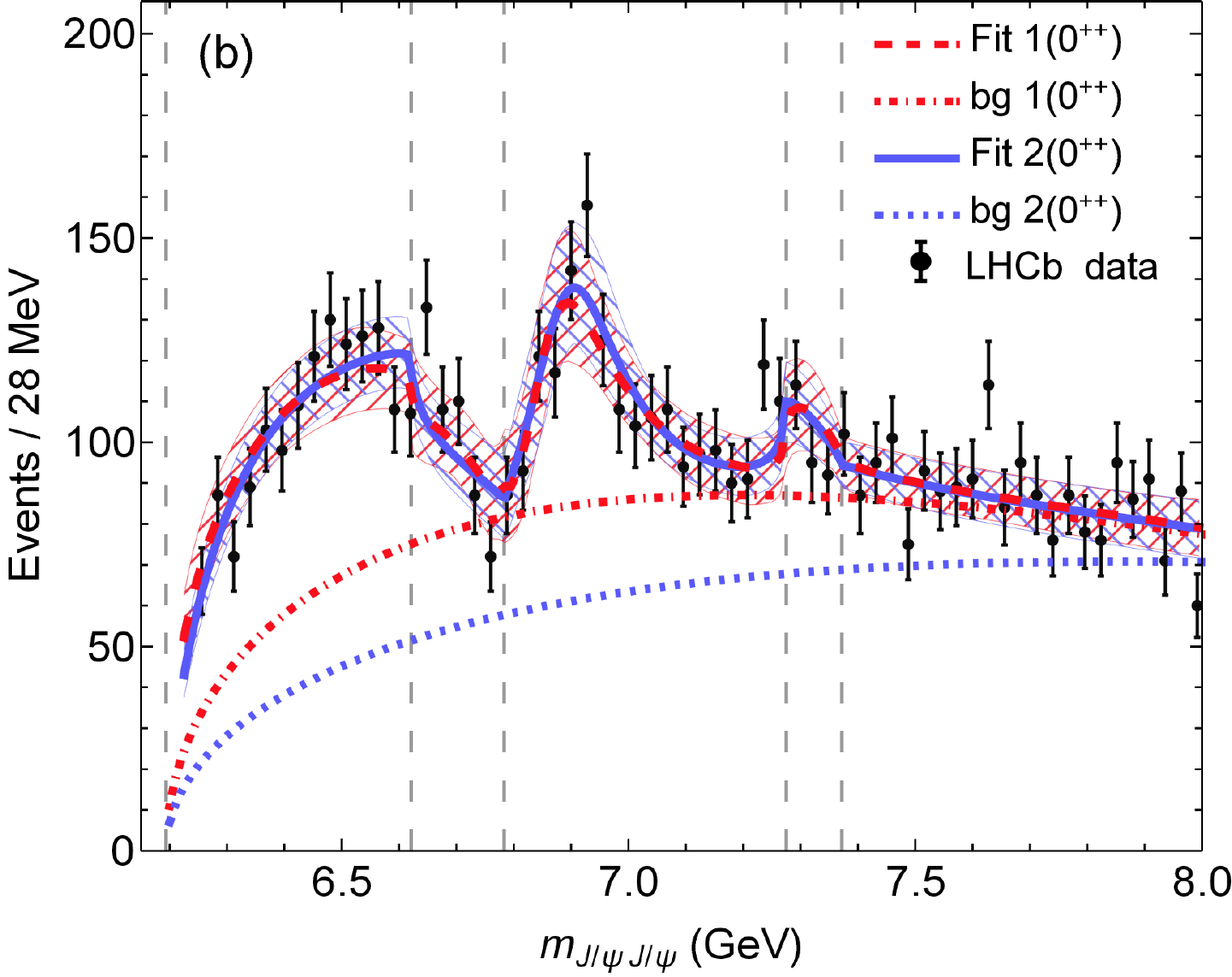}
\includegraphics[scale=0.4]{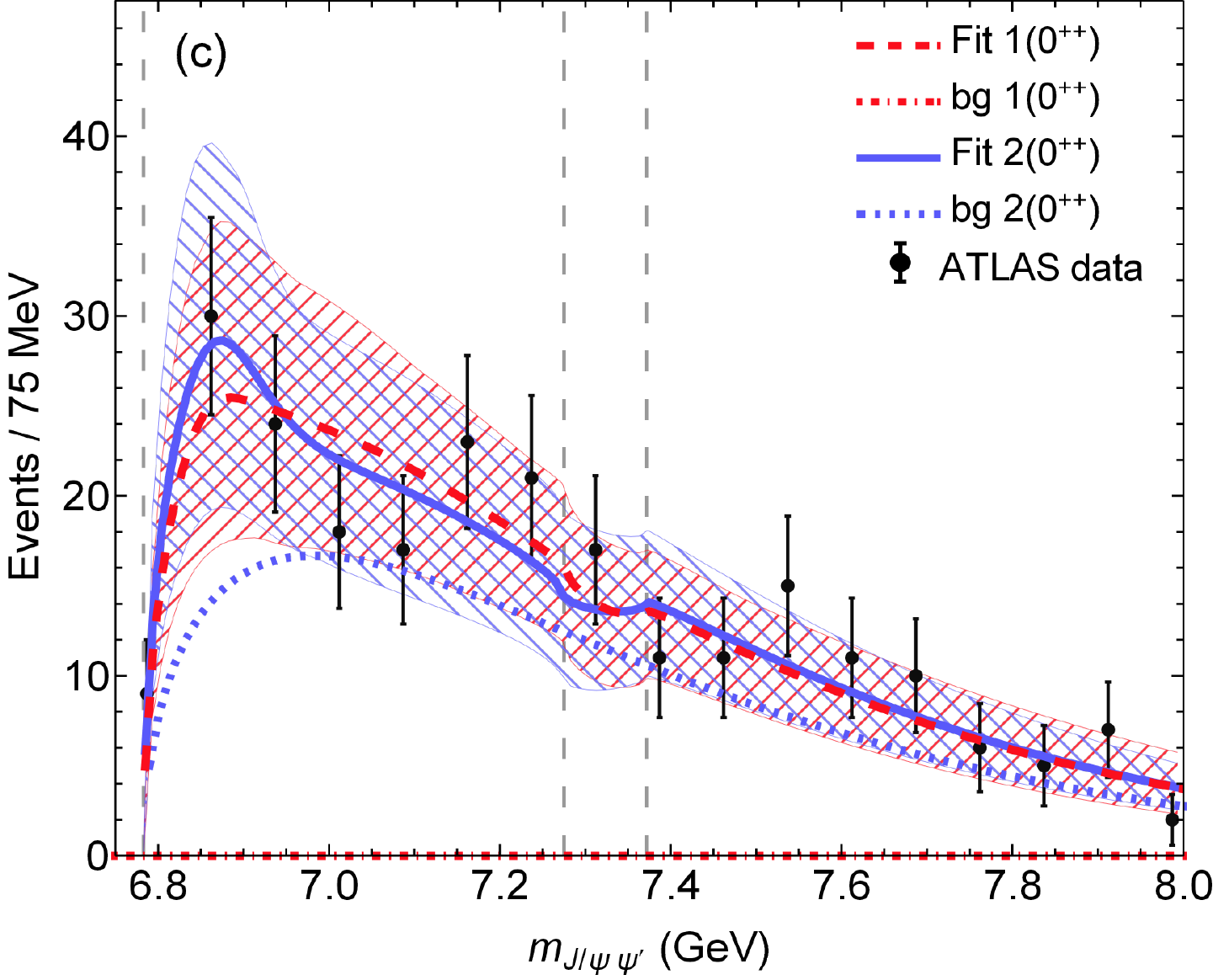} \\
\includegraphics[scale=0.4]{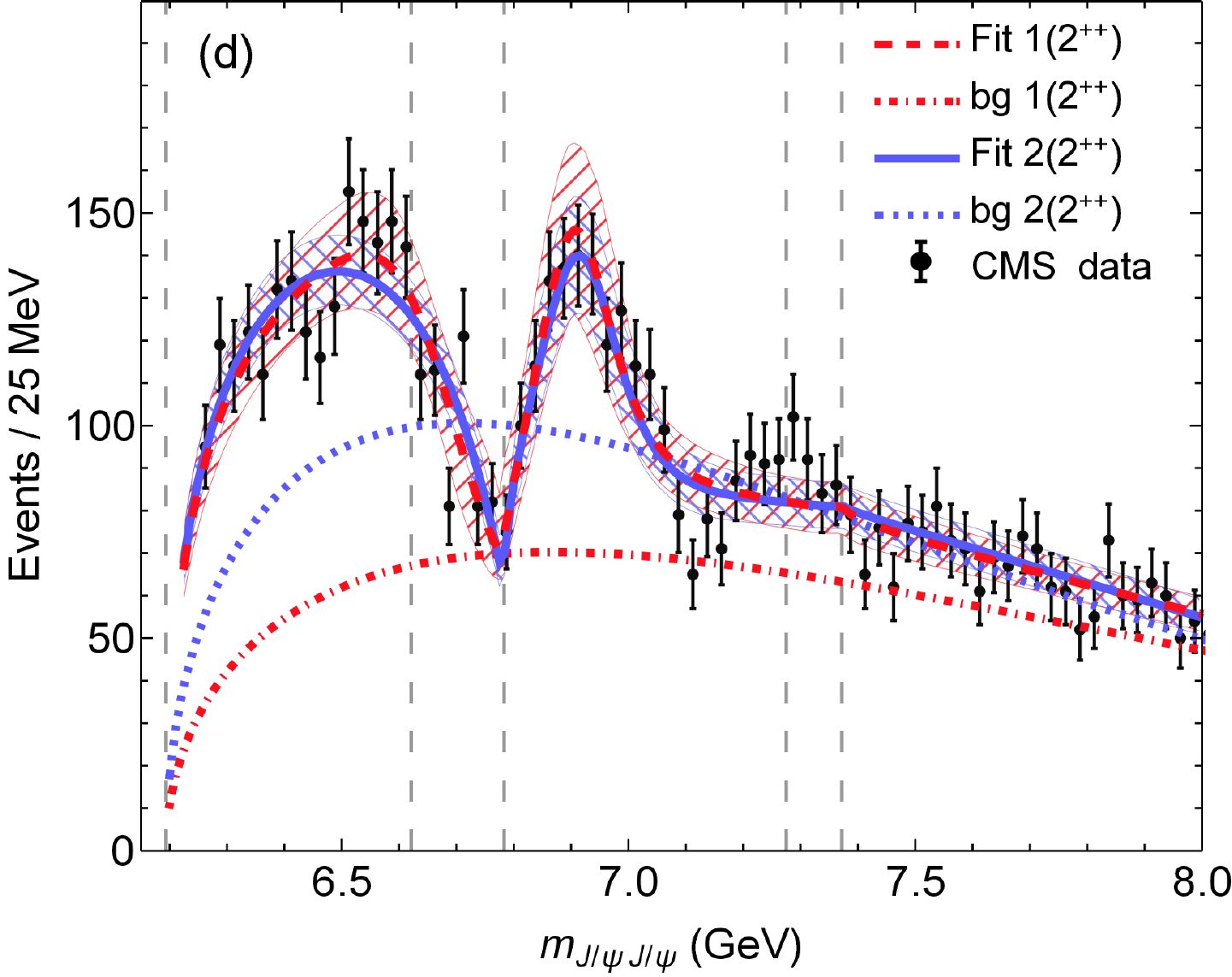}\includegraphics[scale=0.4]{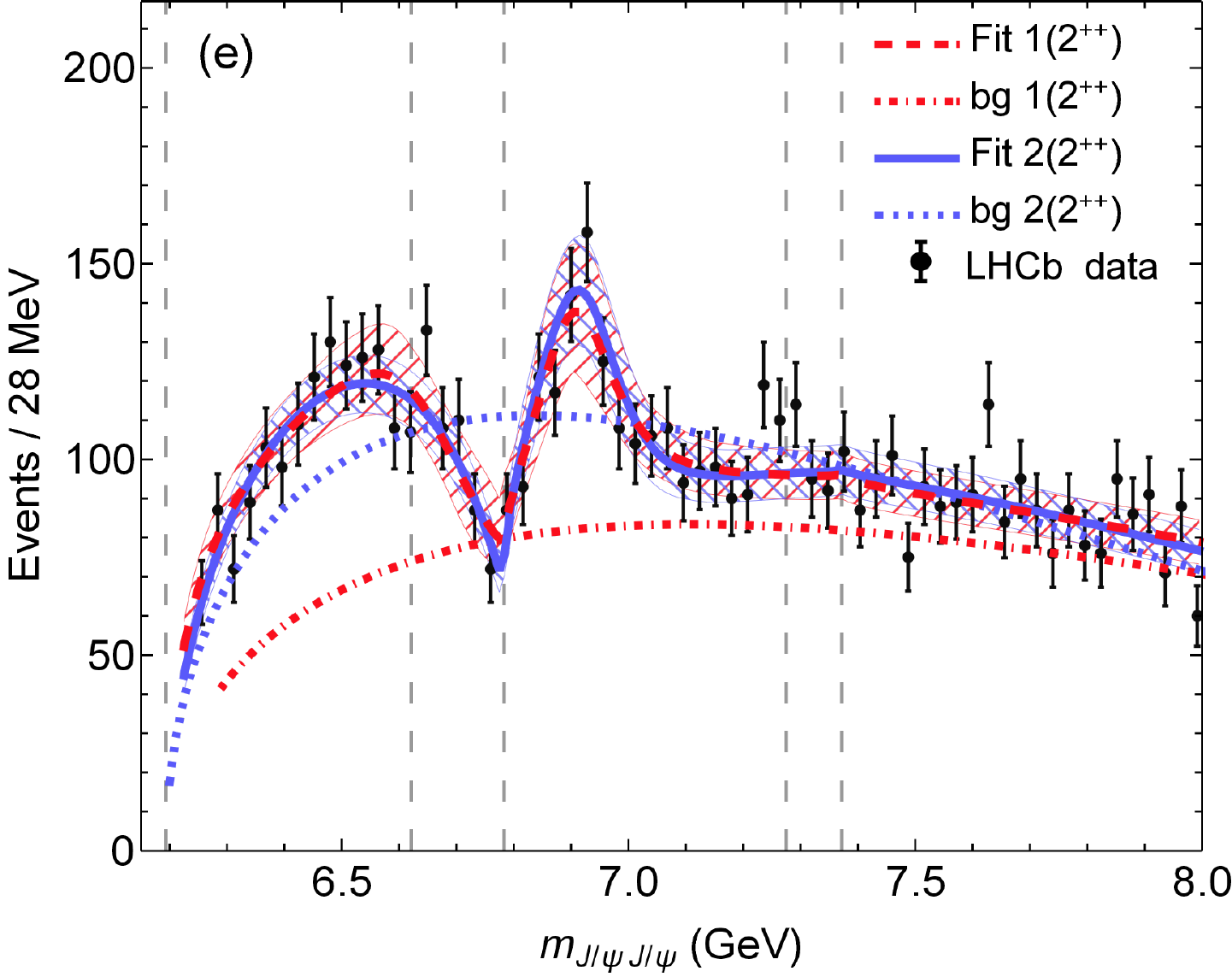}
\includegraphics[scale=0.4]{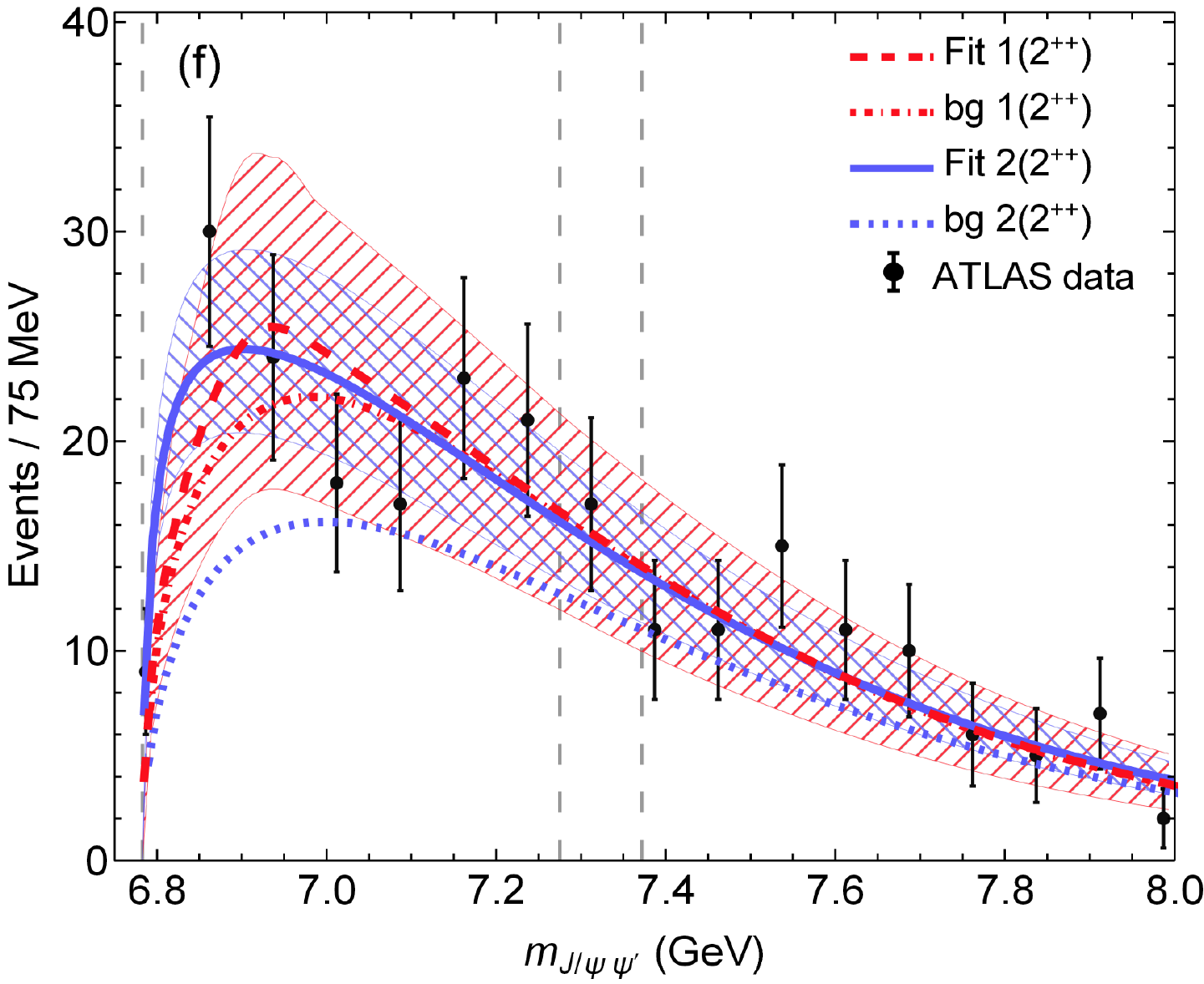}
\caption{(Color online) (a)-(c): The fitted double $J/\psi$ and $J/\psi \psi^\prime$ invariant mass distributions compared with the experimental data for the $J^{PC}=0^{++}$ incoherent (red dashed curves and $\chi^2/\text{d.o.f}=1.08$) and coherent (blue solid curves and $\chi^2/\text{d.o.f}=1.00$) cases. 
(d)-(f): The fitting results of the $J^{PC}=2^{++}$ with $\chi^2/\text{d.o.f}=1.08$ and $\chi^2/\text{d.o.f}=1.09$ for the incoherent and coherent cases, respectively.
The red dot-dashed and blue dotted curves are the corresponding background contributions.
 The error bands (hatched areas) are $1\sigma$ uncertainty propagated from the experimental uncertainties.
 The vertical gray dashed lines are the thresholds of the considered dynamic channels. 
 The experimental data are taken from the LHCb~\cite{LHCb:2020bwg}, CMS~\cite{CMS:2022yhl}, and ATLAS~\cite{ATLAS:2022hhx} collaborations.}
\label{fig:fit01}
\end{figure*}  
 The fitted results are presented in the sub-figures (a)-(c) of Fig.~\ref{fig:fit01}. 
 Both the incoherent and coherent curves can well describe the broader structure
  around $6.5~\gev$ and the narrow structure around $6.9~\gev$. In particular, the peak structure 
  around $7.2~\gev$ in the $J/\psi J/\psi$ channel is also well described. 
However, this structure exhibits itself as a mild dip around $7.2~\gev$ in the $J/\psi\psi^\prime$ channel,
as a result of the unitarity of S-matrix. 
A typical example of similar behavior is the $f_0(980)$ which may exhibit as a peak or dip structure in various amplitude squares~\cite{Ahmed:2020kmp}.
 The reason that the signal of our case is not as significant as that of the $\pi\pi$-$K\bar{K}$ case is the large width.
 This pattern can also happen in the $\eta_c\eta_c$ and $\eta_c\eta_c^\prime$ 
channels, i.e. a peak/dip structure in the $\eta_c\eta_c$/$\eta_c\eta_c^\prime$ channel
between the $\eta_c^\prime\eta_c^\prime$ and $\psi^\prime \psi^\prime$ thresholds. 
This behavior can be used to further confirm the origin of the third peak structure in the $J/\psi J/\psi$ spectrum. 
 \begin{table}[h]
   \caption{The pole positions (in MeV) for
     Scheme-$1$ and Scheme-$2$ of the $J^{PC}=0^{++}$ channel. Here the 1-$\sigma$ statistical uncertainties are presented. The non-interference and interference fit results, i.e. Model I and Model II, of the CMS analysis~\cite{CMS:2022yhl} are listed in the last column for the comparison of our Scheme I and Scheme II, respectively.}
  \begin{tabular}{cccc}
\hline \hline
& poles (MeV)  & notation & CMS data \tabularnewline
\hline\hline
\multirow{3}[0]{*}{Scheme-$1$} 
&$5805^{+126}_{-272}$ & $X^{0^{++}}_1$  & $6552\pm i 62$\\
&$(6849^{+31}_{-39})\pm i(82^{+27}_{-20})$ & $X^{0^{++}}_2$  &$6927\pm i61$ \\
&$(7304^{+8}_{-12})\pm i(135^{+14}_{-12})$ &$X^{0^{++}}_3$ &$7287\pm i47.5$\\
\hline 
\multirow{3}[0]{*}{Scheme-$2$} 
&$5425^{+19}_{-20}$  & $X^{\prime 0^{++}}_1$ & -\\
&$(6067\pm6)\pm i (14\pm 1)$  &$X^{\prime 0^{++}}_2$ &- \\
&$(6883^{+34}_{-45})\pm i (103^{+27}_{-22})$ & $X^{\prime 0^{++}}_3$ & $6736\pm i269.5$\\
&$(7280^{+9}_{-11})\pm i (212^{+48}_{-37})$  & $X^{\prime 0^{++}}_4$ & $6918\pm i93.5$\\
\hline \hline
\end{tabular}
\label{table:pole0}
\end{table}  

Although both the two schemes can describe the data very well, 
their pole structures, identified as bound states/resonances, are different as shown in Tab.~\ref{table:pole0}. 
 Here only poles on the physical Riemann Sheet (RS) and the ones directly connected to the physical region are presented. As the $X^{0^{++}}_1$ is far away from the lowest $\eta_c\eta_c$ threshold
  and located on the unphysical RS as a virtual state, 
  its effect on the physical observables is marginal. 
  Meanwhile, the physical observables have little constraints on this pole.
  That is the reason why the error of this pole is large.  
  The $X^{0^{++}}_2$ is $66^{+31}_{-39}~\mev$
  above the $J/\psi \psi^\prime$ threshold, leaving imprints around $6.9~\gev$ in the $J/\psi J/\psi$ spectrum
and the threshold enhancement in the $J/\psi\psi^\prime$ spectrum. 
   The $X^{0^{++}}_3$ is $29^{+8}_{-12}~\mev$
  above the $\eta_c^\prime\eta_c^\prime$ threshold, leading to the  
significant structures around $7.2~\gev$ in the $J/\psi J/\psi$ spectrum
and the mild dip structure in the $J/\psi\psi^\prime$ spectrum. 
The dip structure around $6.75~\gev$ in the $J/\psi J/\psi$ invariant mass distribution
is because of the inclusion of the dynamical $J/\psi \psi^\prime$ channel.  
For the coherent case, the dip structure around $6.75~\gev$ is sharper than that of the incoherent case.
  The $X^{\prime 0^{++}}_1$ is located on the physical sheet
  with hundreds of MeV below the lowest $\eta_c\eta_c$ threshold. Thus it does not
  deduce a pronounced structure in the lineshapes of the dynamical channels.
  It however may exhibit itself in the inelastic channels,
  for instance, the $J/\psi \mu^+\mu^-$, $\mu^+\mu^-\mu^+\mu^-$ channels, and so on. 
  The $X^{\prime 0^{++}}_2$ is $99\pm6~\mev$
  above the $\eta_c\eta_c$ threshold
  and $127\pm6~\mev$
  below to the $J/\psi J/\psi$ threshold, leading to the threshold increasing behavior in the $J/\psi J/\psi$ lineshape.
   The $X^{\prime 0^{++}}_4$ is $5^{+9}_{-11}~\mev$
  above the $\eta_c^\prime\eta_c^\prime$ threshold and on the RS connecting to the physical one along the positive axis above the $\psi^\prime\psi^\prime$ threshold. It strongly couples to the $\psi^\prime\psi^\prime$ and the $\eta_c^\prime\eta_c^\prime$
  channels, behaving as a peak structure in the $J/\psi J/\psi$ channel and a mild dip structure in the $J/\psi \psi^\prime$
  channel, respectively. 
  %
%\begin{figure*}[th]
%\centering
%\includegraphics[scale=0.4]{pic/21.pdf}\includegraphics[scale=0.4]{pic/22.pdf}
%\includegraphics[scale=0.4]{pic/23.pdf} 
%\caption{The labels are analogous to those of Fig.~\ref{fig:fit01} but with $\chi^2/\text{d.o.f}=1.08$ 
%and $\chi^2/\text{d.o.f}=1.09$ for the $J^{PC}=2^{++}$ incoherent 
%and coherent cases, respectively.}
%\label{fig:fit21}
%\end{figure*} 

It has been shown that the data can be sufficiently well described
by only considering the $0^{++}$ channel. It is not surprising
that the inclusion of the $2^{++}$ channel in addition to the $0^{++}$
does not improve the fit quality significantly
and would lead to uncontrolled uncertainties.
As a result, in what follows we only focus on the $2^{++}$ channel
for the comparison with the $0^{++}$ case. 
For the $2^{++}$ case, the dynamic channels are $J/\psi J/\psi$, $J/\psi \psi^\prime$, $\psi^\prime\psi^\prime$ in the mass order.
 \begin{table}[h]
   \caption{The caption is the same as that of Tab.~\ref{table:pole0} but for
  the $J^{PC}=2^{++}$ channel.}
  \begin{tabular}{cccc}
\hline \hline
& poles (MeV) & notation & CMS data \tabularnewline
\hline\hline
\multirow{3}[0]{*}{Scheme-$1$} 
&$5940^{+58}_{-74}$  &  $X_1^{2^{++}}$  & $6552\pm i62$\\
&$(6677^{+28}_{-49})\pm i(170^{+41}_{-39})$ & $X_2^{2^{++}}$& $6927\pm i61$ \\
&$(6910^{+46}_{-49})\pm i(93^{+16}_{-9})$  & $X_3^{2^{++}}$ &$7287\pm i47.5$ \\
\hline 
\multirow{3}[0]{*}{Scheme-$2$} 
&$6092\pm2$ &    $X_1^{\prime 2^{++}}$  & -  \\
&$(6928\pm6)\pm i (103\pm 2)$  & $X_2^{\prime 2^{++}}$  & $6918+i93.5$\\
\hline \hline
\end{tabular}
\label{table:pole2}
\end{table}   
As shown by the sub-figures (d)-(f) of Fig.~\ref{fig:fit01}, the three lineshapes
can also be well described by the incoherent and coherent schemes
with the reduced chi-square $\chi^2/\text{d.o.f.}$ $1.08$ and $1.09$, respectively.
The dip structure around $6.75~\gev$ also shows
up in the lineshape due to the dynamic $J/\psi \psi^\prime$ channel.
On the contrary, the third peak structure around $7.2~\gev$ does not show up
due to the absence of the 
$\eta_c^\prime\eta_c^\prime$ channel.
Similarly, the dip structure between the $\eta_c^\prime \eta_c^\prime$
and the $\psi^\prime \psi^\prime$ threshold in the $J/\psi \psi^\prime$
channel is also absent. The poles 
denoted as $X_i^{2^{++}}$ and $X_i^{\prime 2^{++}}$ for the $i$th pole for the incoherent and coherent schemes,
with physical impacts
are collected in Tab.~\ref{table:pole2}. 
  The $X^{2^{++}}_1$ is located on the physical sheet and $254^{+58}_{-74}~\mev$
  below the $J/\psi J/\psi$ threshold and strongly couples to this channel.
  It could be viewed as a deeply  $J/\psi J/\psi$ bound state.
  This pole is similar to the predicted $X(6200)$ of Ref.~\cite{Dong:2020nwy}
  and should leave a significant structure in the $\eta_c\eta_c$ channel. 
  The $X^{2^{++}}_2$ has a width of hundreds of MeV,
   leaving it insignificant on the lineshapes. 
The $X^{2^{++}}_3$ is $127^{+46}_{-49}~\mev$
  above the $J/\psi \psi^\prime$ threshold and 
  accounts for the peak structure around $6.9~\gev$ of the $J/\psi J/\psi$ channel
  and the near-threshold enhancement of the $J/\psi \psi^\prime$ channel.  
  The $X^{\prime{2^{ ++}}}_1$ strongly couples to the $J/\psi J/\psi$ channel and is $102\pm 2~\mev$
   below the threshold, making it a $J/\psi J/\psi$ bound state. 
 This pole contributes to the near-threshold enhancement of the $J/\psi J/\psi$ channel.
     The $X^{\prime{2^{ ++}}}_2$ strongly couples to the $\psi^\prime\psi^\prime$ channel
and is located on its physical sheet with $444\pm 6~\mev$   
  below the $\psi^\prime\psi^\prime$ threshold.
  This pole corresponds to the observed $X(6900)$ structure in the $J/\psi J/\psi$
  channel and leads to the near-threshold increasing behavior in the $J/\psi \psi^\prime$ channel. 
\begin{figure}[htbp]
\centering
\includegraphics[scale=0.5]{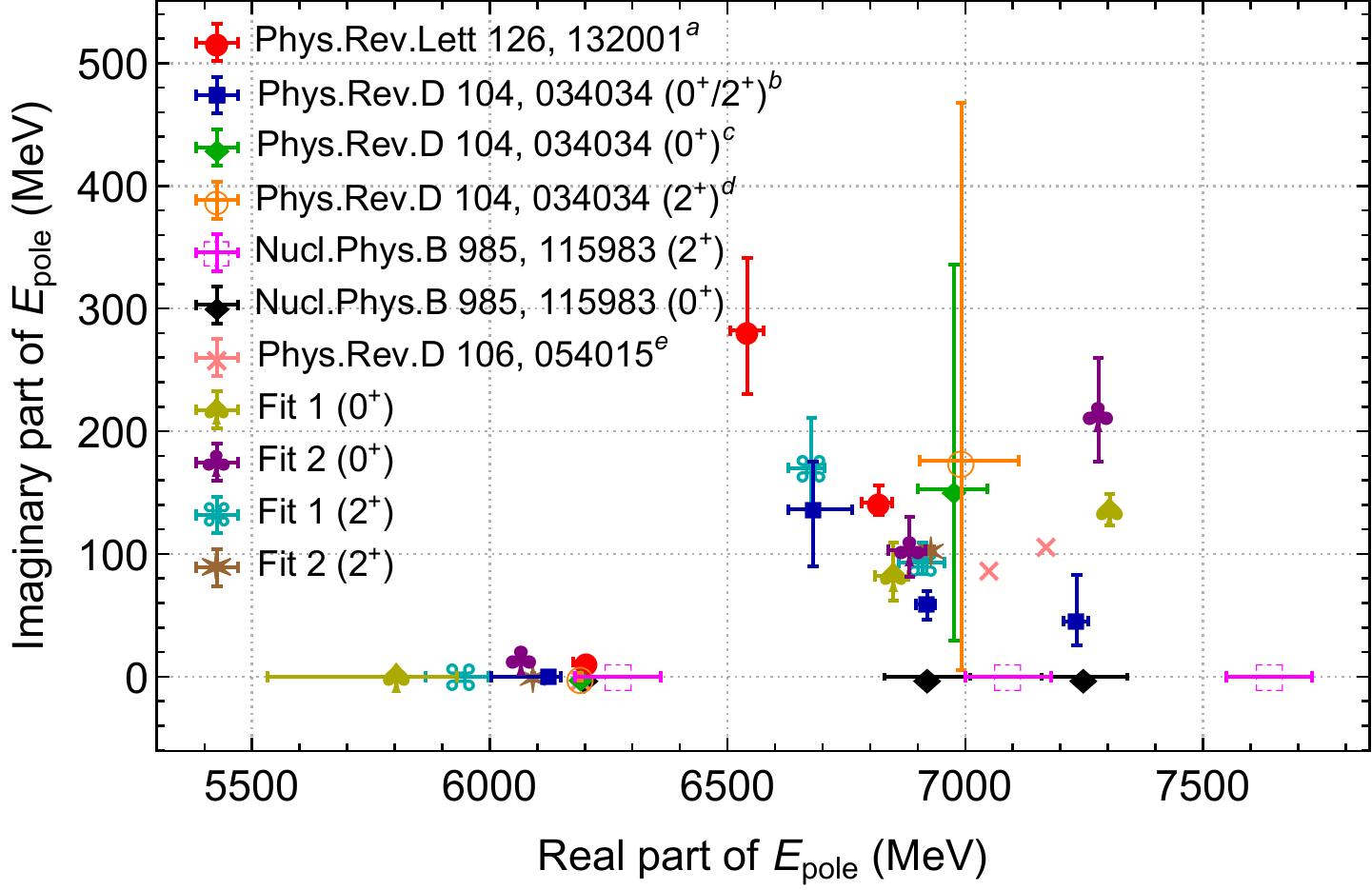}
\caption{(Color online) The comparison of the pole positions around $6.2$ GeV, $6.9$ GeV  and $7.2$ GeV
with those from Refs.~\cite{Dong:2020nwy,Liang:2021fzr,Wang:2022xja,Wang:2022jmb}. 
 Our results are labeled as Fit 1 ($0^+$), Fit 2 ($0^+$), Fit 1 ($2^+$), and Fit 2 ($2^+$). The meaning of  superscripts a-e is given in the context.}
\label{fig:pole}
\end{figure} 

From the above-mentioned four fit schemes, 
it is easy to see that all of them describe the data
well, however, lead to different pole structures.
Especially, in the $2^{++}$ case, neither coherent nor incoherent scheme
can reproduce the third peak. Therefore, once future experiments confirm the third peak,
it should correspond to a $0^{++}$ state. In addition, we also present a comparison 
of the poles with others, e.g. Refs.~\cite{Dong:2020nwy,Liang:2021fzr,Wang:2022xja,Wang:2022jmb}\footnote{  
In Ref.~\cite{Wang:2022xja}, Regge trajectory for ordinary hadrons is employed
for the fully charm system, which has been questioned by Refs.~\cite{Albuquerque:2023rrf,Albuquerque:2020hio}, since the couplings
between the radial excited states and the interpolating currents are
larger than the one of the ground state. }
 around $6.2~\gev$, $6.9~\gev$, $7.2~\gev$ in Fig.~\ref{fig:pole}. 
In the literature, there are various schemes. To distinguish those schemes,
the quantum numbers and the superscripts are labeled in Fig.~\ref{fig:pole}.
The superscript a represents the results of the $J/\psi J/\psi$-$J/\psi \psi^\prime$ coupled channel case of Ref.~\cite{Dong:2020nwy}.
The superscript b represents the results of the $J/\psi J/\psi$-$J/\psi \psi(2S)$-$J/\psi \psi(3770)$-$\psi(2S)\psi(2S)$ coupled channel case with both $0^{++}$ and $2^{++}$ quantum numbers~\cite{Liang:2021fzr}. The superscripts c and d are used to label the above four-channel case with only $0^{++}$ and only $2^{++}$ quantum numbers~\cite{Liang:2021fzr}, respectively.
The superscript e represents Fit-I of Ref.~\cite{Wang:2022jmb} with $\chi_{c1}\eta_c$, $\chi_{c0}\chi_{c1}$, $\chi_{c2}\chi_{c2}$ as the three dynamical channels.
  
In summary, we compare four schemes
 to shed light on the nature of the third peak structure in the $J/\psi J/\psi$
spectrum. As the $\eta_c^\prime\eta_c^\prime$ threshold is very close to the third peak,
one cannot avoid its effect on the structure. 
Since the $0^{++}$ case has the $\eta_c^\prime\eta_c^\prime$
channel as one of the dynamical channels, both the incoherent and coherent cases 
can describe the third peak structure. 
This structure can also express itself as a mild dip structure in the $J/\psi \psi^\prime$
channel between the $\eta_c^\prime\eta_c^\prime$ and $\psi^\prime \psi^\prime$ thresholds.
 On the contrary, the $2^{++}$ case does not include the $\eta_c^\prime\eta_c^\prime$
channel and  
cannot produce the third peak structure in the $J/\psi J/\psi$ channel.
Thus we suggest experimentalists detailed scan both the $J/\psi J/\psi$ and
 the $J/\psi \psi^\prime$ spectra,
especially around $7.2~\gev$ to probe the nature of the third fully charmed state. 
We also find that when the number of the coupled channels increases,
e.g. the six-coupled-channel case for the $0^{++}$ channel,
the predicted $X(6200)$ in Ref.~\cite{Dong:2020nwy} will become a deeper state. 

\vspace{0.3cm}

%\begin{acknowledgement}
{\bf \color{gray}Acknowledgement} We are grateful to Qiang Zhao and Zhi-Hui Guo for the very helpful discussion. P. Y. N is grateful to Ji-Feng Hu for the useful discussion about the data fitting. This work is partly supported by the National Natural Science Foundation of China with Grant No.~12035007, No. 12147128, Guangdong Provincial funding with Grant No.~2019QN01X172, Guangdong Major Project of Basic and Applied Basic Research No.~2020B0301030008. Q.W. is also supported by the NSFC and the Deutsche Forschungsgemeinschaft (DFG, German Research Foundation) through the funds provided to the Sino-German Collaborative Research Center TRR110 ``Symmetries and the Emergence of Structure in QCD" (NSFC Grant No. 12070131001, DFG Project-ID 196253076-TRR 110).

\vspace{0.3cm}

{\bf \color{gray}Author contributions} Peng-Yu Niu and Zhenyu Zhang did the calculations. Meng-Lin Du and Qian Wang drafted the manuscript. All the authors made substantial contributions to the physical and technical discussions and the editing of the manuscript. All the authors have read and approved the final version of the manuscript.

%\bibliography{ref.bib}

\end{document}